\documentclass[prb,aps,floats,twocolumn,showpacs]{revtex4}
\usepackage{amsmath}
\usepackage{graphicx}
\usepackage{epsfig}
\usepackage{float}

\newcommand{\bea}{\begin{eqnarray}}
\newcommand{\eea}{\end{eqnarray}}
\newcommand{\ket}{\rangle}
\newcommand{\bra}{\langle}

\newcommand{\be}{\begin{equation}}
\newcommand{\ee}{\end{equation}}

\begin{document}
 \title{Adiabatic quantum pumping in an Aharonov-Bohm loop and in a Si-like nanowire: Role of interference in real space and in momentum space}
 \author{Sungjun Kim,$^1$ Kunal K. Das,$^{2}$ and Ari Mizel$^1$}
 \affiliation{$^1$ Department of Physics, The Pennsylvania State
                   University, University Park, Pennsylvania 16802, USA    \\
              $^2$ Department of Physics, Fordham University, Bronx, New York 10458, USA}

\begin{abstract}
We study the consequences of interference effects on the current
generated by adiabatic quantum pumping in two distinct
one-dimensional (1D) lattice model. The first model contains an
Aharonov-Bohm (AB) loop within a tight-binding chain of lattice
sites. The static AB phase is shown to strongly affect interference
between the two arms of the loop, serving as an on-off switch and
regulator for the pumped current. The second model simulates pumping
in semiconductors with indirect band-gaps, by utilizing a
tight-binding chain with next-nearest-neighbor coupling. The model exhibits signatures of interference between degenerate conduction band states with different Fermi wavevectors.
\end{abstract}

\pacs{73.23.-b,73.63.-b,72.10.Bg,72.80.Cw}

\maketitle

\section{Introduction}
Adiabatic quantum pumping provides a mechanism to generate a direct
current with no bias\cite{Brouwer}.  By periodically changing the
parameters that define a conduction channel, one forces carriers
down the channel.  The quantity of pumped carriers depends only upon
the path that the parameters take through parameter space and not
upon the speed with which this path is traversed \cite{Altshuler}
(as long as the change is slow enough to remain adiabatic).
Recently, adiabatic quantum pumping has been applied to a number of
different transport contexts, including the generation of spin
polarized
current\cite{Sharma,Mucciolo,Watson,Aono,Wei,Blaauboer,Kim} and
entangled pairs \cite{Beenakker,Samuelsson,Das}.  Experiments
continue to be motivated by sustained interest in this unusual
mesoscopic transport mechanism \cite{Watson, Dicarlo,Buitelaar}.

Since pumping itself is a consequence of quantum interference, it is
of interest to see how it is affected by other quantum
interference phenomena working in conjunction.  In this work we study
two pump configurations which create additional interference effects
due to (i) differences in spatial trajectories, and (ii) due to
competing momenta.  It is natural to think of the former as an
interference effect in position space and the latter as an
interference effect in momentum space.

The first configuration comprises of an Aharonov-Bohm (AB) loop
geometry, subject of many studies associated with AB phase effects
and Fano effects\cite{Gefen,Kobayashi,Aharony,Lu}.  A pair of
parallel quantum dots straddling the linear chain serve as the two
arms of an AB loop. Differences in spatial trajectories are created
in the two arms of the loop. We find surprising sensitivity of the
pumped charge to the presence and magnitude of a \emph{static}
magnetic field associated with the AB effect; even with two time
varying and out of phase parameters, there is no pumped current when
the field is absent.

Early experimental work on adiabatic quantum pumping \cite{Switkes}
observed a pumped current that is symmetric under magnetic field
reversal, while theory predicted no definite symmetry\cite{Aleiner}.
It was suggested that the observed currents may be due to
rectification effects rather than pumping \cite{Piet}. It is thus
desirable to have a model in which pumped current and rectification
current can be definitively distinguished. Rectification currents
should be symmetric under magnetic field reversal, our model is therefore
designed to have a mirror symmetry that causes the pumped charge to
be antisymmetric $q_{pump}(-B) = - q_{pump}(B)$ under magnetic field
reversal. Thus, experiments conducted on a realization of this model
would have some natural explanations of an experimentally observed
current: If the current is anti-symmetric under magnetic field
reversal, it is likely due to quantum pumping, otherwise
unexpected mechanism is involved in generating the resulting
current.

The second model that we consider is a tight-binding chain with
next-nearest-neighbor (nnn) hopping; in that case the minimum of the
conduction band energy need not lie at $k=0$; and the conduction band
can look like that of a semiconductor with an indirect gap. A physical
motivation for this model is the band structure of Si nanowires.
Calculations have predicted an indirect gap \cite{Scheel} when Si
nanowires are grown along certain directions, say, the $[11\bar{2}]$
direction.  The conduction band dispersion relation in our model can
have four Fermi wave vectors at a given Fermi energy $E_{F}$. Our
calculations show rich interference effects between these
wavevectors, including resonant peaks of different heights in the
pumped current.

The rest of the paper is organized as follows. In Sec. II, we
formulate our method for computing the current pumped in a one
dimensional lattice due to an arbitrary localized time varying
potential. In Sec. III, we compute the effects of a static magnetic
field on the AB loop system and demonstrate the spatial interference
effects induced by that field on the pumped current. We study the
effects of nnn coupling in Sec. IV and analyze wavevector interference
phenomena. We summarize our primary conclusions in Sec. V.

\section{Mechanism of Adiabatic Pumping in an one dimensional chain}

The description of adiabatic pumping in mesoscopic systems is based
upon the Landauer-B\"uttiker scattering picture.   For a single
channel in a system describable by a one-dimensional lattice
Hamiltonian, the instantaneous pumped current is \cite{Kim,Entin}
\bea j_{pump}(n) = e \int dE f(E)
            \int_{-\pi}^{\pi} \frac{dp}{2 \pi} \delta(E - E_{p})
            \bra \phi_{p} | j_{n} | \phi_{p} \ket,  \label{current}
\eea
where $j_n$ is the appropriate discrete current operator for the
one-dimensional chain. The system has a single Fermi distribution function
$f(E)$ since pumping operates in the absence of bias. The matrix
elements of the current operator are taken with respect to
scattering states $| \phi_{p}(t) \ket$ of the full
\emph{time-dependent} Hamiltonian $H(t)$ of the system. When the
time dependence of the Hamiltonian is adiabatic, the states $|
\phi_{p}(t) \ket$ can be approximated by an expansion in terms of
the \emph{instantaneous} scattering states $|\chi_{p}(t) \ket$ up to
linear order in the time derivative
 \be | \phi_{p}(t) \ket = | \chi_{p}(t)\ket  - i \hbar G(E_{p}) | \dot{ \chi }_{p}(t) \ket. \label{phieqn}\ee
Here, $G(E_{p})= 1/(E_{p}-H + i \eta)$ is the retarded Green's
function for the \emph{instantaneous} Hamiltonian $H$ with
$\eta=0^{+}$ imposing causality. 
The instantaneous scattering states \be | \chi_{p}(t) \ket = (1+G(E_{p})V) | p \ket,
\ee are exact solutions to the Lippmann-Schwinger equation for the
instantaneous scattering potential $V$ that contains the
time-varying part of the full Hamiltonian $H$ with $| p \ket$ being
a plane wave state. If the time dependence in $H$ is not explicit
but instead arises only through time-dependent parameters, the
states $|\phi_{p}(t)\ket$ and $|\chi_{p}(t) \ket$ inherit time
dependence through those parameters and also do not explicitly
depend on time.

We apply the above considerations to a 1D-lattice Hamiltonian
comprising of the standard time-independent tight-binding
Hamiltonian $H_0$  with nearest-neighbor coupling and a general
time-dependent potential $V(t)$ acting on a finite `scattering
region' of the lattice:
 \bea \label{model-1D-Hamiltonian}
H(t)  &=& H_{0} + V(t), \notag \\
H_{0} &=& -J \sum_{n}(a_{n+1}^{\dag} a_{n} + a_{n}^{\dag} a_{n+1} ), \notag  \\
V(t) &=& \sum_{x,y} V_{xy}(t) a_{x}^{\dag} a_{y}, \eea Here
$a_{n}^{\dag}$ is the electron creation operator at site $n$ and $-J$
is the nearest neighbor hopping coupling strength. The on-site
energy of the sites of the free Hamiltonian $H_0$ is taken to be the
reference energy and hence set to zero. The potential V(t) is
parameterized by the $V_{xy}(t)$ which contribute to the inter-site
coupling strengths $(x\neq y)$ and on-site energy shifts $(x=y)$
within the scattering region.  Since the current is always defined
asymptotically far from the scattering region, $H_0$ determines the
definition of the current operator
\be \label{nearest-current}j_{n} = - \frac{J}{i \hbar} (
a_{n+1}^{\dag} a_{n} - a_{n}^{\dag} a_{n+1} ) \ee
as well as the dispersion relation $E_{p}=-2J \cos p$, where the
spacing between the sites in the chain has been set to unity,
thereby setting a natural length scale.

We evaluate the pumped current for this Hamiltonian in the distant
region $n\rightarrow \infty$ through an adaptation of an analysis we
presented in a recent publication \cite{Kim}. We only consider real
pumping parameters in this paper, so we set
$\dot{V}_{xy}=\dot{V}_{yx}$ for all $x,y$. The pumped current for
the time-dependent Hamiltonian in Eq.~(\ref{model-1D-Hamiltonian})
is then given by
\bea
 \lefteqn{j_{pump}(n) = - \frac{ e J }{\pi}\int dE f(E) \times} \notag \\
 & &  Im \sum_{x,y} \dot{V}_{xy}
      \partial_{E}[ G^{\ast}(E)(n+1,x)  G(E)(n ,y ) ],   \label{derivative}
 \eea
We used the notation $ G(E)(n,x)=\bra 0| a_{n} G(E) a_{x}^{\dag} |0
\ket$.  Since pumping experiments require low temperatures we take
the zero temperature limit whereby $f(E)$ becomes a step function
and an integration by parts transforms Eq. (\ref{derivative}) for
the pumped current to {\small \bea
 j_{pump}(n)
 &=& - \frac{ e J }{\pi}
      Im \sum_{x,y} \dot{V}_{xy}
     G^{\ast}(E_{F})(n+1,x)  G(E_{F})(n ,y ), \label{result}  \notag  \\
 & &
\eea}
where $G(E_{F})(n,x)$ is the full propagator in the energy domain
evaluated at the Fermi energy. We will take our time-varying
parameters to be on-site energies, so only the diagonal elements of
$V_{xy}$ will have non-vanishing time-derivatives and we simplify the
notation by defining $\dot{V}_{xy}= \dot{u}_{x}\delta_{x,y}$.  Using
Dyson's equation\cite{Economou}, $G(E)=G_{0}(E)+ G_{0}(E)V G(E)$,
where $G_{0}(E)=1/( E - H_{0} + i \eta)$, we can show that
$G(E_F)(n,x) = -2 \pi i N(E_F)[ e^{ik_F(n-x)} +\sum_{x^\prime}
u_{x^\prime} e^{ik_F(n-x^\prime)} G(E_F)(x^\prime,x)]$ for $n$ larger
than all $x,x^{\prime}$ in the scattering region.  Thus we have the
identity $G(E_{F})(n+1,x)= e^{ik_{F}} G(E_{F})(n,x)$ for $n$ larger
than all $x$ in the scattering region, and the expression (\ref{result})
for the pumped current becomes
  \bea
  j_{pump}(n)
   = \frac{e}{(2 \pi)^{2}} N( E_{F} )^{-1} \sum_{x} \dot{u}_{x}
     |G(E_{F})(n ,x)|^{2}.  \label{propagator}
  \eea
with $N(E_{F})=\frac{1}{2 \pi} \frac{\partial k }{\partial E_{k} }
\big{|}_{ k=k_{F} }$ being the one-dimensional density of states per
unit length.  In fact, using the Dyson equation result, one can show
that the current is independent of the site index $n$ as one should
expect from charge conservation
  \bea\label{pumped-current}
  j_{pump}
   &=& e N( E_{F} ) \sum_{x} \dot{u}_{x} \times    \\
   & & \big{|} 1 + \sum_{x^{\prime}} u_{x^{\prime}} e^{ -ik_{F}( x^{\prime}-x )} G(E_{F})(x^{\prime},x) \big{|}^{2}.  \label{local}
   \notag\eea
In this expression, the instantaneous pumped current is determined
by the density of states, instantaneous potentials and their time
derivatives, and the {\it local} full propagator (local since $x$
and $x^\prime$ are both in the scattering region where $u_x$ and
$u_{x^\prime}$ are both non-zero), which can be determined from
Dyson's equation from a knowledge of the potential and the free
Green's function.

\section{Pumping Through a Loop Geometry}


\begin{figure}[t]\vspace{-1cm}
\epsfig{figure=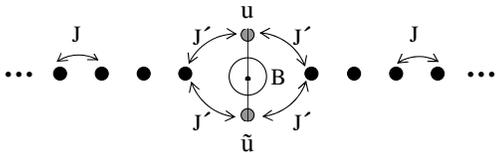,width=1.3\columnwidth}\vspace{-11cm}
\caption{Two quantum dots are connected to left and right leads in
parallel. They form a closed loop in the central region which is
threaded by a magnetic field B. One dot has on-site energy $u$ and
the other has on-site energy $\tilde{u}$. The symbols $-J$ and $-J'$ are
hopping amplitudes. The vertical line shows the plane
of left-right reflection symmetry when $B=0$. \label{Fig:loop}}
\end{figure}

We now apply the considerations of the previous section to determine
the pumped current for our first physical model of interest,
consisting of a 1-D tight-binding chain with a loop in the central
region as shown in Fig.~\ref{Fig:loop}. The Hamiltonian has the
general form shown in Eq~(\ref{model-1D-Hamiltonian}) where
potential is now specifically
\bea
V(t) &=& u(t)b_{0}^{\dag} b_{0} + \tilde{u}(t)\tilde{b}_{0}^{\dag} \tilde{b}_{0}+J[a_{0}^{\dag} a_{-1} +  a_{1}^{\dag} a_{0}]\notag \\
  & & - J^{\prime} e^{ i\varphi / 4 } [ \tilde{b}_{0}^{\dag} a_{-1} + a_{1}^{\dag} \tilde{b}_{0}]
  - J^{\prime} e^{- i\varphi / 4 } [ b_{0}^{\dag} a_{-1} + a_{1}^{\dag} b_{0}]  \notag \\
  & & + ({\rm Hermitean\ conjugate}). \label{Ham}
\eea
To obtain this potential, we imagine displacing site $0$ above the chain so that it forms the upper arm of a loop.  We define the creation operator $b^{\dag}_{0} \equiv a_{0}^{\dag}$, introducing new notation for the displaced state.  We introduce a new site to form the lower arm; the creation operator associated with this new site is $\tilde{b}_{0}^{\dag}$.  Thus, the two parallel sites in the middle of the chain
are represented by $b_{0}^{\dag}$ and $\tilde{b}_{0}^{\dag}$.  These sites
are decoupled from each other but are coupled to the rest of the
chain on either side with a coupling strength of $-J^{\prime}$. The
time-dependence lies at the on-site energies of the two parallel
sites $\tilde{u}$ and $u$, which serve as the pumping parameters.
The two sites straddling the chain create a loop, and a static
magnetic flux $\Phi$ penetrating that loop creates an Aharonov-Bohm
phase difference $\varphi = 2 \pi \Phi / \Phi_{0}$, where $\Phi_{0}=
hc / e$, between the two spatial paths defined by the arms of the
loop, leading to interference effects.

By specifying a cyclic time-dependence of the parameters $u$ and
$\tilde{u}$ and the definition of the potential, we can integrate
the expression in Eq.~(\ref{pumped-current}) to compute the charge
pumped in a full cycle
\bea q_{pump} = \oint dt j_{pump}(n). \eea
We choose a simple square-loop time cycle as shown in Fig. 2(b) in
the space of the two pumping parameters.  The charge pumped over a
cycle is computed numerically and the behavior as a function of the
Aharonov-Bohm phase $\varphi$ is presented in Fig. 2(a). The
dependence as a function of the Fermi vector $k_F$ is shown in
Fig~3.  The results show several interesting features that we now
discuss in detail.


\begin{figure}[t]
\epsfig{figure=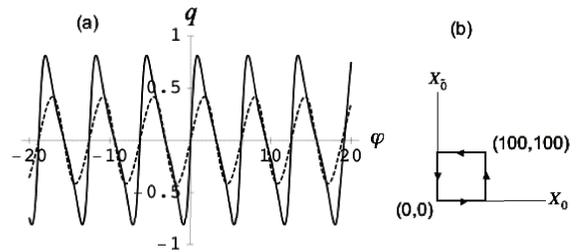,width=3 in} \caption{ (a) Charge $q$ (in
units of $e$ throughout paper) pumped per cycle  vs $\varphi$ (in radians): We set
$J^{\prime}/J=0.5$ and traverse a square-shaped pumping cycle with
corners $(X_0,X_{\tilde{0}}) =(u/J,\tilde{u}/J)= (0,0)$ and
$(100,100)$ as shown in (b).  The solid line in (a) is for
$k_{F}=1.7$, and the dashed line is for $k_{F}=2.1$. This plot shows
the antisymmetry and periodicity of pumped charge as a function of
the magnetic field. The solid line curve near the origin shows that
the pumped current is highly sensitive to small magnetic fields.
\label{Fig:asymmetry}}
\end{figure}

\begin{figure}[t]
\epsfig{figure=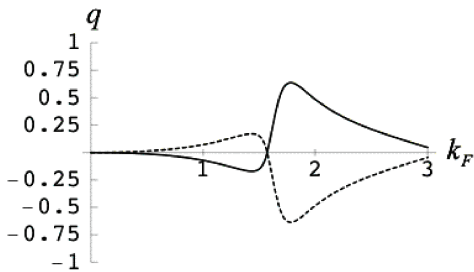,width=2.5 in} \caption{ Pumped charge $q$
vs $k_{F}$: We set $J^{\prime}/J=0.5$ and traverse a square-shaped
pumping cycle with corners $(X_0,X_{\tilde{0}}) = (0,0)$ and
$(100,100)$.  The solid line is for $\varphi= \frac{\pi}{2}$, which
corresponds to $\Phi=\frac{1}{4}\Phi_{0}$, and the dashed line is
for $\varphi= \frac{3\pi}{2}$, which corresponds to
$\Phi=\frac{3}{4}\Phi_{0}$.  Since the current is periodic for
$\varphi$ and is antisymmetric when the sign of $\varphi$ is
reversed, the two curves shown have opposite values of pumped charge
at any given $k_{F}$. \label{Fig:curves}}
\end{figure}

(i) The current vanishes in the absence of the static magnetic field.
The magnetic field plays a crucial role even though it does not vary
in time and therefore never acts as a pumping parameter.  The
necessity for the static magnetic field can be understood by
recognizing that our system has reflection symmetry about a plane
through the two loop sites, orthogonal to the direction of flow. It is
easy to see by simply reversing the magnetic field, and simultaneously
exchanging sites $n$ with $-n$, that the reflection symmetry implies
\bea \oint dt j_{pump}(n,-\varphi) = -\oint dt j_{pump}(n,\varphi)
\label{symmetry}\eea
This is an example of a discrete symmetry \cite{Aleiner} which
causes the pumped charge to be antisymmetric under reversal of
magnetic field.  This antisymmetry is also manifest in Fig.
\ref{Fig:curves}, where the two traces of the pumped charge, as a
function of the Fermi vector $k_F$, are exactly antisymmetric
because one is for $\varphi=\pi/2$ and the other is for
$\varphi=3\pi/2\equiv -\pi/2$.

More intuitively, when the magnetic field is absent, the two parallel
quantum dots effectively pump at the same location along the current
flow.  The two potentials combine into a single parameter, and a
single parameter cannot pump any current.  The static magnetic field
breaks the reflection symmetry, and creates a phase shift between the
two dots.  They no longer lie symmetrically at the same location in
the current flow, and now act as two separate pumping parameters,
producing a current.  In Ref.  \onlinecite{Shin-Hong}, a related effect
was seen, in a very different configuration involving two loops where
the pumping parameters were time-varying magnetic fields.  In that
work, the static field also breaks the symmetry, allowing a pumping current to arise.


(ii) There is a periodicity in the charge pumped with respect to the
phase introduced by the static magnetic field, as shown in Fig.
\ref{Fig:asymmetry}.  This periodicity is an expected feature of AB
phenomena.  The periodicity of $\varphi=2\pi$ causes the pumped charge
to vanish when the phase difference due to the magnetic field between
the two arms of the loop vanishes, equivalent to having no field at
all.

(iii) The pumped charge also vanishes for certain values of $k_F$
specifically at $k_F=n\pi/2$.  This can be seen in Fig.
\ref{Fig:curves}, where the pumped charge is plotted as a function
of $k_F$ for two different values of the Aharonov-Bohm phase
$\varphi$. The reason is that an electron picks up a phase of $\pm
k_F$ as it moves from site to adjacent site. Therefore in traversing
one of the arms of the loop it picks up a total phase of $\pm
2k_F\pm\varphi/2$, and in going through the other arm, the
corresponding total phase change is $\mp 2k_F\mp\varphi/2$.  When
$k_F=\pi/2$, they become respectively $\pm \pi\pm\varphi/2$ and $\mp
\pi\mp \varphi/2\equiv 2\pi-(\mp \pi\pm\varphi/2) \equiv
\pm\pi\pm\varphi/2$. This shows that the phases accumulated in both
paths are identical and therefore the symmetry-breaking effect of
the magnetic field is annulled and the pumped charge vanishes as if
the magnetic field was not there at all.

We conclude the discussion of this system by noting that the pumped
current is quite sensitive to changes in the external magnetic field
as seen from the solid curve in Fig. \ref{Fig:asymmetry}(a).  This
can provide a way of precisely controlling the magnitude and
direction of the pumped current without changing the time-varying
pumping parameters in any way. In addition the fact that the absence
of the field leads to vanishing current suggests an obvious
application as a switching mechanism.

\section{Pumping on a chain with next-nearest-neighbor hopping}
\begin{figure}[b]
\epsfig{figure=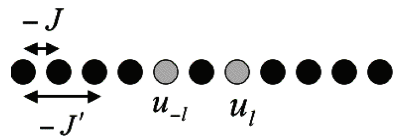,width=2 in} \caption{ We consider a 1D
tight-binding chain with both nearest neighbor and
next-nearest-neighbor coupling, where $-J$ is the strength for
nearest neighbor hopping, and $-J^{\prime}$ is the strength for
next-nearest-neighbor hopping. On this chain, we assume only two
sites (gray dots) located at $-l$ and $l$ have non-zero on-site
energies $u_{-l},u_{l}$. \label{Fig:nnn}}
\end{figure}

We now turn our attention to the pumped current on a chain with
next-nearest-neighbor (nnn) hopping (see Fig. \ref{Fig:nnn}).   We
consider the following Hamiltonian \bea
H(t) &=& H_{0} + V(t),  \notag \\
H_{0} &=& -J \sum_{n} (a^{\dag}_{n+1} a_{n} + a^{\dag}_{n} a_{n+1})
      \notag\\ &&-J^{\prime} \sum_{n} (a^{\dag}_{n+2} a_{n} + a^{\dag}_{n} a_{n+2}), \notag  \\
V(t) &=& u_{-l}(t) n_{-l} + u_{l}(t) n_{l}, \eea
where $- J^{\prime}$ is the nnn hopping amplitude and
$n_{l}=a^{\dag}_{l}a_{l}$ is the number operator on site $l$. The
on-site energy is taken to be zero for all sites except for the
sites $\pm l$.  The energies at those sites $u_l$ and $u_{-l}$ are
the time varying pumping parameters.

\begin{figure}[b]
\epsfig{figure=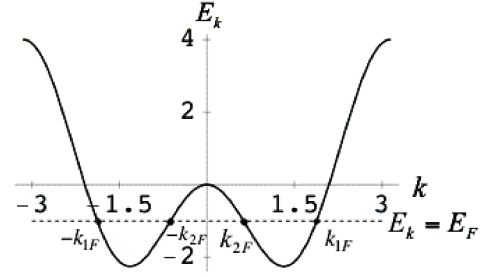,width=2.5 in} \caption{ The dispersion
relation for $J=1$ and $J^{\prime}=-1$. It has a double-well shape.
The dashed line is for $E_{k}=E_{F}=-1$;  it has four crossing
points with the doublewell curve, corresponding to four Fermi wave
vectors $\{ k_{1F},k_{2F},-k_{1F},-k_{2F} \}$.
\label{Fig:doublewell}}
\end{figure}

The dispersion relation for this model (taking the lattice constant
to be unity) is $E_{k}=-2J \cos k - 2 J^{\prime} \cos 2 k$. We
assume a positive $J$ and negative $J^{\prime}$ so that the
dispersion relation yields a double-well shape as shown in Fig.
\ref{Fig:doublewell}. This ``indirect gap'' shape is physically
relevant, being reminiscent, for instance, of the band structure of
certain Si nanowires \cite{Scheel}.  For any total energy $E < 0$ 
in Fig. \ref{Fig:doublewell}, solving $E_{k}=E$ yields four solutions $\{ \pm k_{1}(E),\pm
k_{2}(E)\}$, related by $\cos k_{1}+\cos k_{2}=-J/(2J')$. At zero
temperature the pumping dynamics is determined by the Fermi energy
$E=E_F$; we denote the corresponding wavevectors $\{ \pm k_{1F},\pm
k_{2F}\}$.  We choose the convention that $k_{1}> k_{2} > 0$.


We need to extend the analysis of Sec. II to get the pumped current
by defining an appropriate discrete current operator for this
extended chain. Since there is a nnn hopping process, we define the
current operator using the continuity equation
$\partial_{t}[\rho(n)+ \rho(n+1)]+ J(n+1) - J(n-1)=0$, where
$\rho(n) = \bra \psi | a^\dagger_n a_n | \psi \ket$ and $J(n) = \bra
\psi | J_{n} | \psi \ket$. We are led to the definition of the
operator {\small \bea J_{n} &=& - \frac{J}{i \hbar} ( a^{\dag}_{n+1}
a_{n} - a^{\dag}_{n} a_{n+1} )
     - \frac{J^{\prime} }{i \hbar} ( a^{\dag}_{n+1} a_{n-1} - a^{\dag}_{n-1} a_{n+1} )  \notag \\
     & & - \frac{J^{\prime} }{i \hbar} ( a^{\dag}_{n+2} a_{n} - a^{\dag}_{n} a_{n+2} ).
\eea}

We compute the pumped current using this current operator in the
expression for the current (\ref{current}), instead of $j_{n}$; (No
confusion should arise between the current operator $J_n$ and the
tunneling parameters $J$ and $J^\prime$.) In the zero temperature
limit and at points far away ($n \rightarrow \infty$) from the
action of the pumping potential, the pumped current is
{\small \bea J_{pump} (n) &=& e \int dE f(E)
           \int_{-\pi}^{\pi} \frac{dp}{2 \pi}
                     \delta( E-E_{p}) \bra \phi_{p} | J_{n}| \phi_{p} \ket   \label{current-double-well} \\
&=& - \frac{eJ}{\pi} Im \sum_{x=\pm l} \dot{ u }_{x}
       G(E_{F})(n,x) G^{\ast}(E_{F})(n+1,x)   \notag   \\
& &  - \frac{eJ^{\prime}}{\pi} Im \sum_{x=\pm l} \dot{ u }_{x}
       G(E_{F})(n-1,x) G^{\ast}(E_{F})(n+1,x)  \notag   \\
& & - \frac{eJ^{\prime}}{\pi} Im \sum_{x=\pm l} \dot{ u }_{x}
       G(E_{F})(n,x) G^{\ast}(E_{F})(n+2,x).       \notag
\eea}


\begin{figure}
\epsfig{figure=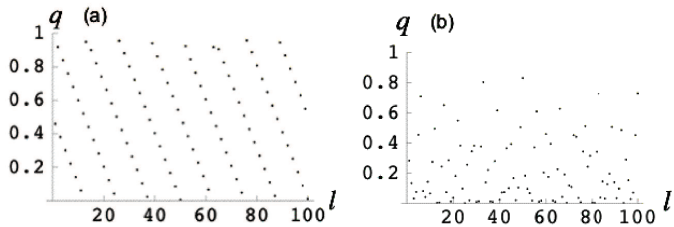,width=3.5 in} \caption{ Pumped charge $q$
vs $l$ (there is exactly one value of $q$ for each choice of $l$ in
both (a) and (b)).  In both (a) and (b), the Fermi level is
$E_{F}=-1.5$, and the pumping cycle is square shaped with lower-left
corner $(0,0)$ and upper-right corner $(100,100)$. The distance
between the two time-dependent potentials is $2l$.    (a) Standard
chain with $J=1$ and no nnn hopping, $J^{\prime}=0$.  A regular
periodicity is evident. (b) Chain with nnn hopping, $J=1$ and
$J^{\prime}=-1$.  The dependence of $q$ on $l$ is quite irregular
with no obvious pattern. \label{Fig:comparison}}
\end{figure}


Some care is needed in applying Eq.~(\ref{current-double-well}).
Because of the shape of the dispersion relation, the velocity $v_{k}=
\frac{1}{\hbar} \frac{\partial E_{k}}{\partial k}$ for $+k_{2}$ is
negative while the velocity for $-k_{2}$ is positive.  Thus incoming
waves from the left reservoir correspond to wave vectors $\{
k_{1},-k_{2} \}$ rather than $\{ k_{1},k_{2} \}$. To keep this
straight, when computing the full Green's function $G(E)=1/(E-H+i
\eta)$, we use $\eta = \eta_{1}= 0^{+}$ at the $k_1$ singularity in
the denominator and $\eta = \eta_{2}= 0^{-}$ at the $k_2$ singularity
in the denominator.  As a result, left incoming waves with $k_{1}$ or
$-k_{2}$ have transmitted waves with the correct physical wave vectors
$\{ k_{1},-k_{2}\}$ and reflected waves with the correct physical wave
vectors $\{-k_{1},k_{2}\}$.

Using Dyson's equation to compute the full Green's function, we
derive an explicit expression for the instantaneous pumped current

\begin{widetext}
\bea J_{pump} = - \frac{2eJ^{\prime} }{\pi} \sum_{x=\pm l} \dot{ u
}_{x}
      \bigl[  g_{1}^{2}
   | d(-x) +  e^{2ik_{1}x} h(-x) |^{2}
    \sin k_{1} ( \cos k_{1}  - \cos k_{2} )   \notag  \\
     + g_{2}^{2}
   | d(-x) +  e^{-2ik_{2}x} h(-x) |^{2}
   \sin k_{2} ( \cos k_{1}  - \cos k_{2} ) \bigr] \big{|}_{E=E_{F}},
\eea

\end{widetext}
where
 {\small
 \begin{subequations}
 \bea
 d(x) &=& \frac{1 - u_{x} G_{0}(E)(0,0)}{Z_{x}}, \\
 h(x) &=& \frac{u_{x} G_{0}(E)(x,-x)}{Z_{x}}, \\
 Z_{x} &=& [ 1 - u_{x} G_{0}(E)(0,0) ] [ 1 - u_{-x} G_{0}(E)(0,0) ]  \notag  \\
       & &  - u_{-x} u_{x} G_{0}^{2}(E)(x,-x),  \label{denominator}  \\
 g_{1} &=&  \frac{1}{ 2 i J \sin k_{1} + 4 i J^{\prime} \sin 2 k_{1}  },  \\
 g_{2} &=&  - \frac{1}{ 2 i J \sin k_{2} + 4 i J^{\prime} \sin 2 k_{2}  },   \\
 G_{0}(E)(x,y) &=& g_{1} e^{ik_{1}|x-y|} + g_{2} e^{-ik_{2}|x-y|}.
 \eea\end{subequations}}

The free Green's function $G_{0}(E)(x,y)$ is found to be the sum of
two terms $g_{1} e^{ik_{1}|x-y|}$ and $g_{2} e^{-ik_{2}|x-y|}$. As a
result, interference effects arise between the two wave vectors $\{
k_{1},-k_{2} \}$.  In particular, the $G_{0}^{2}(x,-x)$ term in
$Z_{x}$, defined above, contains a factor of the form
$e^{2i(k_{1}-k_{2})|x|}$ which depends on the sum of the two wave
vectors, $k_{1}-k_{2}$ and the distance $2 |x|$ between the
locations of the two time varying sites $n_l$ and $n_{-l}$ in the
potential $V(t)$. This factor of $e^{2i(k_{1}-k_{2})|x|}$ admits an
interpretation in terms of interference between the wavefunction
with vector $k_{1}$ and the one with vector $-k_{2}$.  Physically
speaking, the dependence of the pumped current on the separation
$2|x|= 2 l$ between the two points of action of the potentials in
$V$ depends not only on $k_{1}$ and $-k_{2}$ individually but also
on $k_{1}-k_{2}$. This leads to a rather
irregular pattern for the pumped current as function of the
separation $2l$ as seen in Fig.~\ref{Fig:comparison}(b).  This is
quite distinct from the regular pattern, shown in
Fig.~\ref{Fig:comparison}(a)  for comparison, for standard nearest
neighbor coupling i.e. when $J'=0$ .  In all cases $u_l$ and
$u_{-l}$ trace out a square-shaped time cycle like the one shown in
Fig. \ref{Fig:asymmetry}


\begin{figure}[t]
\epsfig{figure=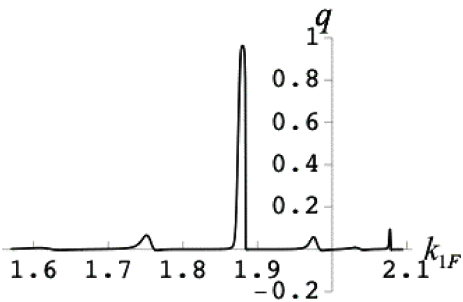,width=2.5 in} \caption{ Pumped charge $q$
vs $k_{1F}$: We set $J=1$, $J^{\prime}=-1$ and $l=10$, and use a
square-shaped pumping cycle with left-lower corner at $(4,4)$ and
right-upper corner at $(100,100)$. High peak near $k_{1F}=1.9$ is
related to resonance transmission and low peak between $k_{1F}=1.7$
and $k_{1F}=1.8$ indicates destructive interference between two wave
vectors. \label{Fig:resonance}}
\end{figure}


In Fig. \ref{Fig:resonance}, we investigate how the pumped current
depends upon the Fermi wave vector $k_{1F}$; note that $k_{1F}$
determines the Fermi energy $E_F$, via the dispersion relation, as
well as $-k_{2F}$. The high peak near $k_{1F}=1.9$ results from a
resonance effect, as we now explain.  Note that the transmission
will be greatest when the denominator $Z_{l}$ in Eq.
(\ref{denominator}) is as small as possible. For large $u_{l},
u_{-l}$, this means minimizing the terms in $Z_l$ that are
multiplied by the product $u_{l} u_{-l}$. This leads to the
condition $G_{0}^{2}(E)(l,-l)-G_{0}^{2}(E)(0,0)= 0$, which
corresponds to three resonance conditions for the wave vectors:
$e^{4ik_{1F}l}=1$, $e^{-4ik_{2F}l}=1$, and
$e^{2i(k_{1F}-k_{2F})l}=1$.  Now when the Fermi wave vector is close to $k_{1F}\simeq 1.9$, where the strong peak occurs in
Fig. \ref{Fig:resonance}, all three conditions are satisfied and a large
pumped charge $q\simeq 1$ arises. In case of the smaller peak
between $k_{1F}=1.7$ and $k_{1F}=1.8$ in Fig.
\ref{Fig:resonance}, the value of $k_{1F}$ satisfies the first two
resonance conditions but not the third.
Physically, this can be interpreted
as destructive interference between the two wave vectors,
resulting in a smaller pumped charge.

\bigskip

\section{Conclusion}

We have used two distinct physically relevant models to illustrate how
the charge pumped in an adiabatic pumping process can be strongly
influenced by interference effects in both position space and momentum
space.  Specifically, we have demonstrated that pumped current can be
generated using a single Aharonov-Bohm loop provided a static magnetic
field is present.  Although that field itself is not an active pumping
parameter, the pumped charge is very sensitive to its magnitude and
direction, a feature that can used for very delicate control of the
pumping process. As we noted, the fact that the pumped current is
antisymmetric in the magnetic field suggests an experimental means of
clearly distinguishing pumped current and current arising in another
way that lacks this antisymmetry.  Using a
next-nearest-neighbor-coupling lattice model to simulate the
indirect-gap present in the band-structure of Si, we demonstrated that
interference effects in momentum space cause strong and weak resonant
peaks in the pumped current per cycle.  The models considered here demonstrate that interference phenomena provide powerful techniques for altering adiabatic pumping behavior.

\acknowledgments

We gratefully acknowledge the support of the Packard Foundation,
NSF NIRT program Grant No. DMR-0103068 and the Research
Corporation.


\end{document}